\documentclass[acmsmall]{acmart}

\AtBeginDocument{%
  }

\usepackage{diagbox}
\usepackage{multirow}
\usepackage{booktabs}
\usepackage{slashbox}
\usepackage{tcolorbox}
\usepackage[table]{xcolor}
\usepackage{xspace}
\usepackage{wrapfig}
\usepackage{float}
\usepackage{graphicx}
\usepackage{algorithm}
\usepackage{algpseudocode}
\usepackage{hyperref}
\usepackage{enumitem}
\newcommand{\toolname}{{\textsc{ReasonVul}}\xspace}
\newcommand{\mycode}{\ttfamily}
\usepackage{xcolor}

\begin{document}

\title[ReasonVul: A Multi-perspective Reasoning Framework for Enhanced Vulnerability Detection]{Three Heads Are Better Than One: A Multi-perspective Reasoning Framework for Enhanced Vulnerability Detection}
\author{Xin Peng}
\orcid{0000-0001-8642-8582}
\affiliation{%
  \institution{National University of Defense Technology}
  \city{Changsha}
  \country{China}
}
\email{xinpeng@nudt.edu.cn}

\author{Bo Lin}
\orcid{0000-0001-5905-4677}
\affiliation{%
  \institution{National University of Defense Technology}
  \city{Changsha}
  \country{China}
}
\email{linbo19@nudt.edu.cn}

\author{Jing Wang}
\orcid{0000-0001-6370-3086}
\affiliation{%
  \institution{National University of Defense Technology}
  \city{Changsha}
  \country{China}
}
\email{wangjing@nudt.edu.cn}

\author{Xiaoling Li}
\orcid{0000-0002-9479-2541}
\affiliation{%
  \institution{National University of Defense Technology}
  \city{Changsha}
  \country{China}
}
\email{lixiaoling@nudt.edu.cn}

\author{Jun Ma}
\orcid{0000-0003-2258-0854}
\affiliation{%
  \institution{National University of Defense Technology}
  \city{Changsha}
  \country{China}
}
\email{majun@nudt.edu.cn}

\author{Jie Yu}
\orcid{0009-0007-1545-7010}
\affiliation{%
  \institution{National University of Defense Technology}
  \city{Changsha}
  \country{China}
}
\email{yj@nudt.edu.cn}

\author{Xiaoguang Mao}
\orcid{0000-0003-4204-7424}
\affiliation{%
  \institution{National University of Defense Technology}
  \city{Changsha}
  \country{China}
}
\email{xgmao@nudt.edu.cn}

\author{Shangwen Wang}
\authornote{Shangwen Wang is the corresponding author.}
\orcid{0000-0003-1469-2063}
\affiliation{%
  \institution{National University of Defense Technology}
  \city{Changsha}
  \country{China}
}
\email{wangshangwen13@nudt.edu.cn}

\renewcommand{\shortauthors}{X. Peng, B. Lin, j. Wang, X. Li, J. Ma, J. Yu, X. Mao, S. Wang}

\begin{abstract}
Automated vulnerability detection is crucial for enhancing software security by identifying potential flaws that attackers could exploit, thereby reducing the reliance on labor-intensive manual code audits. Recent advancements have shifted towards leveraging large language models (LLMs) for vulnerability detection, with techniques like Vul-RAG and VulnSage demonstrating progress through structured prompting and external knowledge integration. However, these approaches typically rely on a single reasoning paradigm, limiting their ability to address the complex and diverse nature of real-world vulnerabilities. To overcome these limitations, we propose \toolname, a novel multi-perspective reasoning framework that harnesses cognitive synergy among three specialized LLM agents, each embodying a distinct reasoning mode. The framework begins with independent analyses of the source code, followed by a structured debate mechanism to resolve conflicts through iterative rebuttal and revision, ultimately converging on a collaborative judgment. Evaluated on the PrimeVul dataset, \toolname achieves a PairAcc of 40.00\% and an F1-score of 72.52\%, surpassing the best baseline by 81.24\% in PairAcc. Further tests on the JITVUL dataset confirm its generalizability, with a PairAcc of 28.67\%. Additionally, we analyzed 542 conflict cases and found that 389 were correctly resolved, highlighting the framework's ability to uncover hidden vulnerabilities through the error-correction mechanism driven by the debate. This work emphasizes the importance of multi-perspective reasoning and collaborative validation in achieving robust and comprehensive vulnerability detection in real-world software systems.
\end{abstract}





\maketitle

\section{Introduction}
Software vulnerabilities, as a long-standing and fundamental challenge in the realms of software engineering and security, have become increasingly severe with the rapid expansion of system scale and complexity~\cite{lin2020software,peng2025keep,cao2021bgnn4vd}. Attackers can exploit these vulnerabilities to gain unauthorized access or steal sensitive data, thereby severely threatening the security and reliability of systems. As a traditional security measure, manual code auditing is a labor-intensive process that requires significant time investment and relies heavily on expert knowledge, making it unsuitable for the fast-paced modern development cycle~\cite{zhuang2021smart, cheng2021deepwukong, cao2022mvd,wang2025divide}. Therefore, automated vulnerability detection (AVD) has gradually become an important research direction in both academia and industry, aiming to reduce the burden of manual review by automatically identifying potential vulnerabilities~\cite{li2018vuldeepecker}.

Over the past decade, AVD techniques have primarily included program analysis-based approaches ~\cite{arusoaie2017comparison} and deep learning-based approaches~\cite{steenhoek2023empirical}. Program analysis methods, such as static analysis and symbolic execution, uncover potential vulnerabilities by examining program behavior. While theoretically capable of offering strong coverage and interpretability, these methods are often hindered by challenges such as path explosion and complex control flows, which result in low efficiency and poor scalability when applied to large-scale projects~\cite{lipp2022empirical, batur2021novel}. In contrast, deep learning approaches leverage neural networks to extract feature patterns from historical vulnerability data, demonstrating certain advantages in large-scale scenarios~\cite{harzevili2023survey}. However, these methods are highly sensitive to the quality of training datasets, and their generalization ability is limited when encountering previously unseen vulnerabilities~\cite{jain2023code,yang2023does}. 

Recently, large language models (LLMs) pre-trained on massive amounts of source code and security knowledge bases have opened up a new research paradigm for vulnerability detection~\cite{ridoy2024enstack}. With their exceptional code semantic understanding and contextual reasoning capabilities, LLMs can imitate human experts and identify vulnerabilities that are difficult to detect using traditional methods. Preliminary research~\cite{fu2023chatgpt} shows that with direct zero-shot prompts, LLMs can achieve performance comparable to deep learning-based methods. Moreover, researchers~\cite{sheng2025llms, nong2024chain,li2025everything} have even explored more sophisticated prompting strategies, such as techniques like Chain of Thought (CoT), to achieve accurate vulnerability detection.
Basically, vulnerability detection can be viewed as a process of reasoning. Given a piece of code, the goal is not merely to recognize surface-level patterns or syntactic anomalies, but to infer deeper semantic relationships, potential execution paths, and the presence of unsafe behaviors under specific conditions. This inherently involves a form of reasoning about the program's logic and behavior, whether it is symbolic reasoning, statistical reasoning, or learned reasoning. Motivated by this perspective, we propose to categorize existing vulnerability detection approaches based on the underlying form of reasoning they employ into three fundamental reasoning modes~\cite{okoli2023inductive}~\footnote{Most other reasoning types can be regarded as variations or combinations of these three modes in cognitive psychology.}:

\begin{itemize}[leftmargin=*]
    \item \textbf{Deductive Reasoning.} This top-down process derives specific conclusions from general principles. In AVD, deductive methods involve applying predefined security rules or standards, such as CERT C~\cite{seacord2008cert}, to identify violations in code snippets. For example, approaches like SAVul~\cite{SAVul} and approaches described in IRIS~\cite{li2024iris} analyze code by checking for deviations from established secure coding practices, such as unsanitized user inputs leading to SQL injection vulnerabilities. These methods excel at detecting well-defined, rule-based vulnerabilities but may struggle with context-specific issues not covered by existing rules.

    \item \textbf{Inductive Reasoning.} This bottom-up process forms general principles from specific observations. The inductive methods analyze historical vulnerability data or known exploits to identify recurring patterns, such as the misuse of specific APIs, and apply these patterns to detect similar vulnerabilities in new code. Existing approaches~\cite{yu2024retrieval, liao2025klrag}, such as Vul-RAG~\cite{Vul-rag}, leverage retrieval-augmented generation (RAG) to match code against known vulnerability patterns, making them effective for detecting common, recurring vulnerabilities but less adept at identifying novel or context-dependent issues.

    \item \textbf{Abductive Reasoning.} This process seeks the most plausible explanation for an observation by reasoning backward from effect to cause. The abductive methods hypothesize potential exploit scenarios, such as authentication bypass or race conditions, and trace back to identify code weaknesses that could enable such outcomes. For instance, VulnSage~\cite{VulnSage} adopts a reflective perspective to analyze code by reasoning from observed outcomes. These approaches excel in detecting complex, context-specific vulnerabilities that require an understanding of intricate program behaviors. However, they may fail to identify vulnerabilities that deviate from the assumed scenarios, as their effectiveness largely depends on the quality of the hypothesized attack chains.
\end{itemize}

Despite the promise of individual approaches, we posit that their confinement to a single reasoning mode constitutes a critical limitation~\cite{he2024idea}.
This narrow focus overlooks the complementary strengths of different cognitive strategies, thereby hindering their effectiveness against the multifaceted nature of real-world vulnerabilities~\cite{sheng2025evaluating}. 
The integration of multiple reasoning paradigms has historically addressed complex problems. A notable case is Louis Pasteur's refutation of the theory of spontaneous generation.~\footnote{\href{https://bio.libretexts.org/Bookshelves/Microbiology/Microbiology_(Boundless)/01\%3A_Introduction_to_Microbiology/1.01\%3A_Introduction_to_Microbiology/1.1C\%3A_Pasteur_and_Spontaneous_Generation}{https://bio.libretexts.org}
}
His experiments relied on inductive reasoning, as he generalized from observations of microbial growth; deductive reasoning, as he designed the swan-neck flask experiment to test clear predictions; and abductive reasoning, as he inferred that contamination was best explained by microbes carried in the air.
This case illustrates how diverse modes of scientific reasoning can complement one another to yield robust conclusions. Inspired by this, we hypothesize that integrating multiple forms of reasoning could similarly enhance vulnerability detection.
To investigate this hypothesis, we conducted an empirical study on the PairVul dataset~\cite{Vul-rag} (detailed in Section~\ref{sec:motivation}), applying three reasoning paradigms to the same set of programs. Our results reveal that the sets of vulnerabilities detected by each mode are largely distinct, with significant complementarity among them.
These findings suggest that \textit{merely relying on a single reasoning paradigm is insufficient for effectively detecting the complex vulnerabilities present in real-world scenarios}.

Motivated by this observation, we propose \toolname, a collaborative framework designed to harness the power of \textbf{multi-perspective reasoning} through a process of \textbf{cognitive synergy}.
The core idea is to assemble experts embodying diverse reasoning paradigms, enabling multi-perspective structured analysis and cross-validation of viewpoints to enhance both the accuracy and reliability of the detection. Specifically, we design three agent roles that simulate security experts specializing in deductive, inductive, and abductive reasoning. During the detection process, the three agents first conduct independent structured reasoning on the source code and produce intermediate conclusions accompanied by explanatory justifications. When their conclusions diverge, a debate stage is initiated. In this phase, the three expert agents critically examine one another’s conclusions, assess the alignment between their own reasoning and that of others, and, in cases of disagreement, choose either to rebut or to revise their prior conclusions, ultimately converging on a collaborative inference grounded in multi-perspective.

To assess the effectiveness of \toolname, we conducted extensive experiments on the PrimeVul~\cite{PrimeVul} dataset, utilizing pairwise accuracy (PairAcc)~\cite{yildiz2025benchmarking} as a key evaluation metric. PairAcc is a strict metric that considers a vulnerability-fix pair correctly classified only if both the vulnerable code and its fixed version are accurately identified. Our evaluation compares \toolname against eight state-of-the-art baselines, including learning-based (LineVul~\cite{linevul}, Coca~\cite{coca}, MoEVD~\cite{MoEVD}) and LLM-based (SAVul~\cite{SAVul}, Vul-RAG~\cite{Vul-rag}, VulnSage~\cite{VulnSage}, GPTLens~\cite{GPTLens}, and VulTrial~\cite{VulTrial}) methods. The results demonstrate that \toolname significantly outperforms all baselines, achieving a PairAcc of 40.00\% and an F1-score of 72.52\%, surpassing the best baseline (VulTrial) by 81.24\% in PairAcc and 29.09\% in F1-score. Furthermore, we conducted an in-depth analysis of the debate mechanism, revealing that it correctly resolves about 72\% of conflicting cases (389 out of 542), compared to only about 9\% with majority voting. These findings highlight \toolname’s superior capability to detect complex vulnerabilities through cognitive diversity and structured debate.

In summary, our study makes the following contributions:

\begin{itemize}[leftmargin=*]
    \item \textbf{New Perspective.} We are the first to integrate deductive, inductive, and abductive reasoning within a multi-agent collaborative framework for vulnerability detection, thereby establishing a paradigm that more closely mirrors the cognitive processes of human experts.

    \item \textbf{Approach.} We design and implement \toolname, which leverages LLMs to assume the roles of security experts with distinct reasoning strengths, and employs a debate mechanism to drive their collaborative decision-making.

    \item \textbf{Experiment.} We conduct extensive experiments on the PrimeVul dataset, and the results demonstrate that \toolname significantly outperforms other state-of-the-art LLM-based approaches. The code and results are publicly available on our online repository.
\end{itemize}

\section{Related Works}
\subsection{Learning-Based Vulnerability Detection Approaches}
The general paradigm of these approaches is to train deep learning models on labeled code snippet datasets to automatically learn deep features and patterns of vulnerabilities~\cite{zhang2023detecting, wu2022vulcnn}. They are categorized into graph-based and token-based methods based on code representation~\cite{zhou2025large}. Specifically, graph-based methods focus on capturing the structural information of code. They model code as graph structures and leverage graph neural networks (GNNs) to learn representations~\cite{zhang2024dshgt, peng2024multi, peng2023dual}. For instance, Devign~\cite{zhou2019devign} constructs a code property graph by integrating abstract syntax trees (AST), control flow graphs (CFG), and data flow graphs (DFG) to capture complex structural relationships, utilizing a gated graph neural network (GGNN) as an encoder to learn vulnerability-related features. While graph-based models excel at capturing explicit code structures, the emergence of Transformer architectures and large-scale pretraining has opened new avenues for understanding deep code semantics~\cite{zhao2024coding}. Consequently, token-based methods leverage pretrained models like CodeBERT to extract code semantics. For example, studies such as LineVul~\cite{linevul}, Stagedvulbert~\cite{jiang2024stagedvulbert}, and Vulcobert~\cite{xia2024vulcobert} have adopted CodeBERT~\cite{feng2020codebert} as their core feature extractor, achieving rapid convergence and superior performance. To further enhance detection accuracy, the MoEVD~\cite{MoEVD} model introduces a Mixture-of-Experts framework, decomposing the broad vulnerability detection task into subtasks targeting specific CWEs, thereby enabling finer-grained classification and detection.
Despite the advantages of learning-based methods, they also face certain limitations. First, their performance is highly dependent on the quality and diversity of training data, and their generalization ability is often limited when encountering vulnerabilities not present in the training set~\cite{nguyen2024deep}. Second, these methods primarily focus on syntactic or structural patterns in code, making it challenging to identify vulnerabilities requiring contextual understanding at the semantic or logical level~\cite{risse2025top}. These shortcomings have spurred exploration into methods based on LLMs.

\subsection{LLM-Based Vulnerability Detection Approaches}
Unlike learning-based methods, LLMs leverage their natural language understanding and code comprehension capabilities to identify vulnerabilities without extensive task-specific fine-tuning. Preliminary explorations show that LLMs exhibit considerable potential in vulnerability detection even under zero-shot or few-shot settings~\cite{zhang2024prompt}. Recent studies further reveal the deeper capabilities of LLMs. For instance, it has been found that when guided by structured prompting techniques such as CoT, LLMs can outperform pre-trained language models like CodeBERT that have been specifically fine-tuned~\cite{zibaeirad2024vulnllmeval,gao2023far, PrimeVul}. This finding is significant as it suggests that the advanced reasoning abilities unlocked through prompting may be more effective for handling complex vulnerability analysis tasks than the pattern-recognition abilities gained through traditional supervised learning. To further enhance the depth of LLM reasoning, researchers have explored more sophisticated mechanisms inspired by human expert cognition~\cite{steenhoek2024err,zhou2024large}. These include the adoption of self-reflection mechanisms, which prompt LLMs to critically review and revise their initial conclusions, as well as the use of RAG, which incorporates relevant historical vulnerability patterns from external knowledge bases into the reasoning process to improve accuracy~\cite{daneshvar2025vulscriber}.

Recently, researchers have constructed adversarial detection frameworks through role-playing~\cite{wei2025advanced,wen2024evalsva}. The core idea of such methods is to simulate expert teams engaging in mutual rebuttal to enhance detection performance. For example, VulTrial~\cite{VulTrial} adopts a courtroom-inspired design to build a more robust vulnerability detection framework, where LLMs play the roles of prosecutor, defender, and judge, followed by another LLM acting as a jury to make the final verdict on the presence of a vulnerability. Although these methods demonstrate the potential of collaborative reasoning, their role design is fundamentally based on role-playing. As their interactions largely amount to mutual questioning, the reasoning process approaches a form of extended self-reflection~\cite{huynh2025detecting}. In contrast, \toolname proposes a fundamentally different approach. Instead of mimicking procedural roles, our agent framework designates each agent to embody a core cognitive reasoning mode (i.e., deduction, induction, or abduction). This design seeks to build a more comprehensive analysis process by integrating distinct yet complementary logical perspectives, thereby more faithfully replicating the multidimensional reasoning process of human experts when tackling complex problems.

\section{Motivation}
\label{sec:motivation}

\begin{wrapfigure}[15]{r}{0.4\linewidth}
  \centering
  \includegraphics[width=\linewidth]{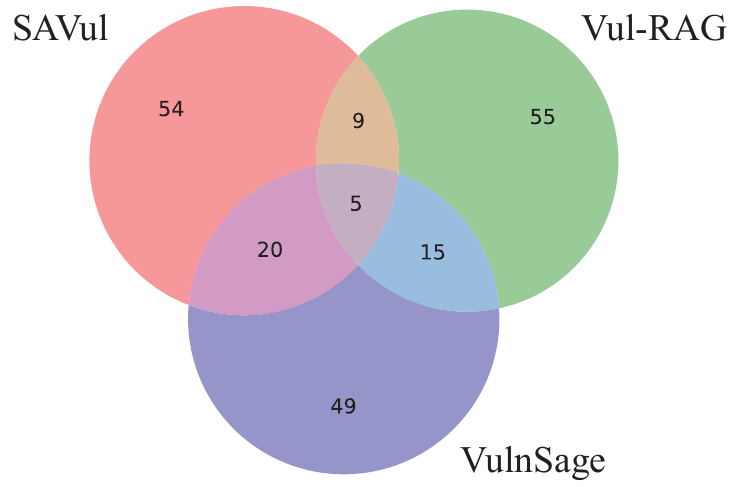}
  \caption{The overlap of vulnerabilities correctly identified by SAVul (Deductive), Vul-RAG (Inductive), and VulnSage (Abductive) on the PairVul dataset.}
  \label{fig:motivation-vven}
\end{wrapfigure}

While recent LLM-based methods have shown considerable promise in vulnerability detection, their effectiveness is often constrained by an implicit reliance on a single reasoning paradigm. To investigate this limitation, we conducted an empirical study on the PairVul dataset~\cite{Vul-rag}, analyzing how representative state-of-the-art tools, each embodying a distinct reasoning mode, perform on real-world vulnerabilities. We selected SAVul~\cite{SAVul} as an exemplar of deductive reasoning, Vul-RAG~\cite{Vul-rag} for inductive reasoning, and VulnSage~\cite{VulnSage} for abductive reasoning. Our quantitative and qualitative analyses, detailed below, reveal that these paradigms exhibit unique strengths and critical blind spots, underscoring the need for an integrated approach.

We first evaluated the performance of SAVul, Vul-RAG, and VulnSage on the PairVul dataset and visualized the overlap of vulnerabilities they correctly identified. As shown in the Venn diagram in Fig.~\ref{fig:motivation-vven}, the intersection among the sets of detected vulnerabilities is strikingly small. Each method identifies a large number of unique vulnerabilities that the other two miss, demonstrating that deductive, inductive, and abductive reasoning paradigms are highly complementary rather than redundant. This quantitative evidence confirms that relying on a single reasoning paradigm inherently leaves significant portions of vulnerabilities undetected due to paradigm-specific blind spots.

To further illustrate this complementarity, we selected three representative cases from the PairVul dataset, each highlighting the distinct strengths and limitations of the reasoning paradigms. These cases, drawn from the Linux kernel, demonstrate how deductive, inductive, and abductive reasoning excel at detecting different types of vulnerabilities.

\begin{figure*}[!t]
\centering
\includegraphics[width=0.99\textwidth]{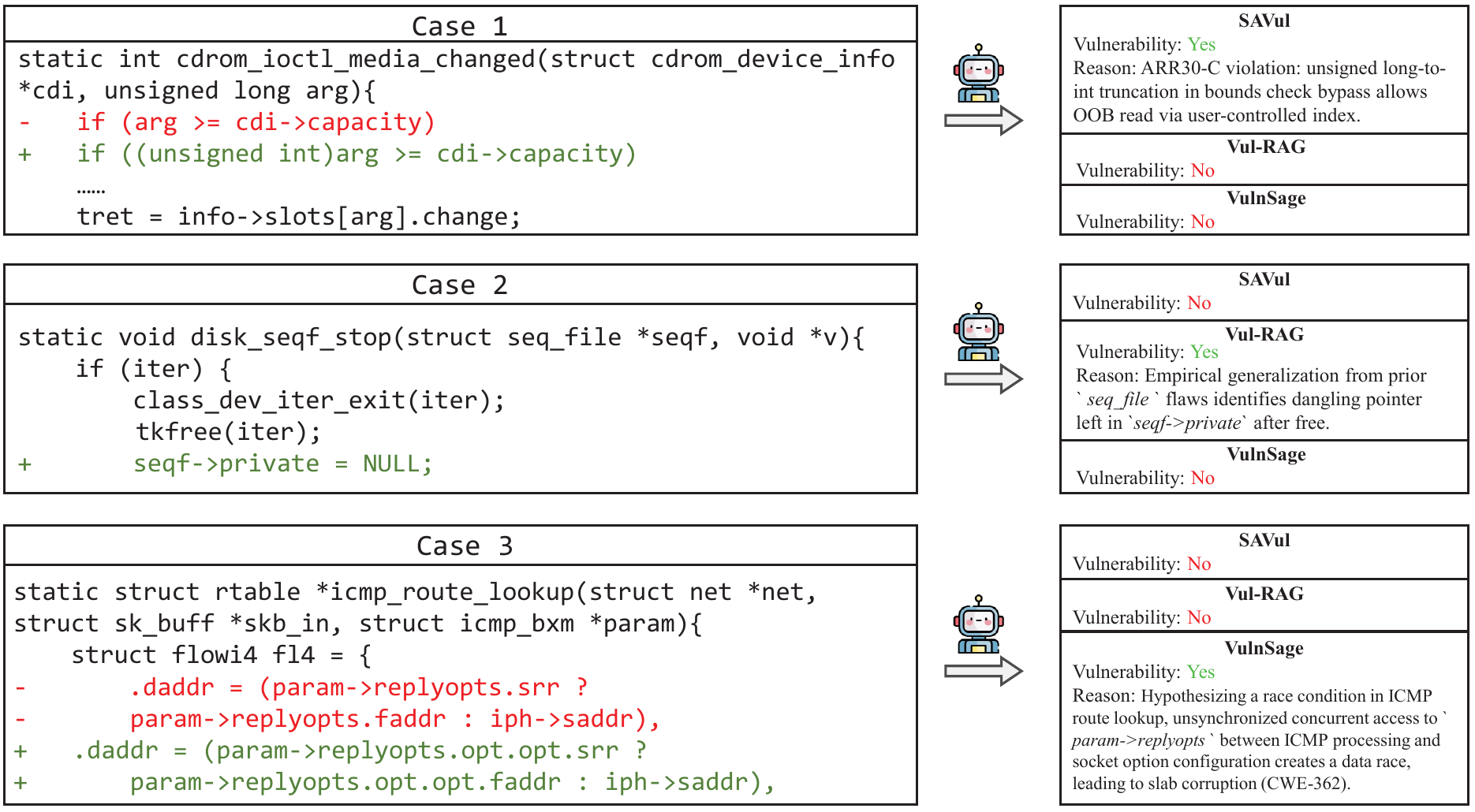}
\caption{Three motivating cases from the Linux kernel, selected from the PairVul dataset, demonstrating the distinct roles of deductive, inductive, and abductive reasoning in vulnerability detection.}
\label{fig:motivation-case}
\end{figure*}

\textbf{Case 1: Integer Overflow.}
As illustrated in Fig.~\ref{fig:motivation-case} (Case 1), an integer overflow vulnerability exists in the {\mycode cdrom\_ioctl\_media\_changed} function. The code casts an {\mycode unsigned long} variable {\mycode arg} to an {\mycode unsigned int} before a bounds check. This explicit downcasting can cause a large input value for {\mycode arg} to wrap around to a small integer, thereby bypassing the {\mycode arg < cdi->capacity} check and leading to an out-of-bounds access at {\mycode info->slots[arg]}.
Among the three approaches, only SAVul, leveraging deductive reasoning, successfully detected this vulnerability. Its top-down approach applied the general security principle that ``type casting between integer types of different sizes can cause overflows'' directly to the code, thereby flagging the {\mycode (unsigned int)arg} cast as a violation.
In contrast, Vul-RAG, relying on inductive reasoning, failed to detect the issue due to its inability to match the code to known vulnerability patterns in its knowledge base, limiting its generalization capability. Similarly, VulnSage, using abductive reasoning, incorrectly assumed the issue was a simple out-of-bounds access. It traced back to the bounds check, deemed it sufficient, and missed the subtle integer overflow within the type cast, concluding no vulnerability existed.

\textbf{Case 2: Use-After-Free.} Case 2 involves a use-after-free vulnerability in the {\mycode {disk\_seqf\_stop}} function. The code frees the memory allocated to the {\mycode {iter}} pointer using {\mycode {kfree(iter)}} but neglects to set the {\mycode {seqf->private}} pointer, which references the freed {\mycode {iter}}, to {\mycode {NULL}}. Subsequent calls to the function under specific conditions may dereference this dangling pointer, causing memory corruption or undefined behavior. 
This vulnerability was successfully detected only by Vul-RAG. Its pattern-based approach retrieved a similar vulnerability pattern related to the {\mycode  {seq\_file}} function from its knowledge base, identifying the dangling pointer in {\mycode {seqf->private}} as a match. 
However, SAVul and VulnSage failed to detect the issue. SAVul could not link the missing nullification to a specific security rule, while VulnSage’s hypothesized attack scenario did not account for the pointer impact, missing the vulnerability entirely. 

\textbf{Case 3: Race Condition.} The third case, selected from the PairVul dataset, involves a complex race condition in the Internet Control Message Protocol (ICMP) routing logic, which handles network communication tasks such as error reporting and diagnostics in the Linux kernel. Improper access to a shared structure under specific network traffic conditions can cause slab corruption and system crashes. Both SAVul and Vul-RAG, relying on deductive and inductive reasoning, respectively, misclassified the vulnerability as non-existent. 
SAVul struggles to extract general security principles suitable for deductive application from this code, resulting in the race condition being overlooked, and Vul-RAG could not generalize from its knowledge base due to the rarity of similar cases. 
In contrast, VulnSage, using abductive reasoning, successfully detected the vulnerability. By hypothesizing a race condition in the ICMP route lookup, it traced back to unsynchronized access to {\mycode {param->replyopts}} between ICMP processing and socket option configuration, correctly identifying a data race leading to slab corruption (CWE-362).

These case studies confirm that deductive reasoning excels at detecting rule-based vulnerabilities, inductive reasoning is effective for pattern-based, common vulnerabilities, and abductive reasoning is suited for uncovering complex, context-dependent vulnerabilities. Our findings reveal a fundamental limitation in LLM-based vulnerability detection: reliance on a single mode of reasoning. Consequently, merely adopting any one of these paradigms is insufficient to effectively address the diverse and complex nature of real-world vulnerabilities. Motivated by this insight, we propose \toolname, a collaborative framework designed to harness the power of \textbf{multi-perspective reasoning} through a process of \textbf{cognitive synergy}. The core idea of \toolname is to first allow agents specializing in deductive, inductive, and abductive reasoning to independently analyze the code, maximizing the breadth of initial findings. Subsequently, it initiates a structured debate mechanism where agents cross-validate, challenge, and synthesize their findings, aiming to converge on a final judgment with significantly higher accuracy and reliability.

\section{Approach}
\subsection{Overview}
\begin{figure*}[!t]
\centering
\includegraphics[width=0.99\textwidth]{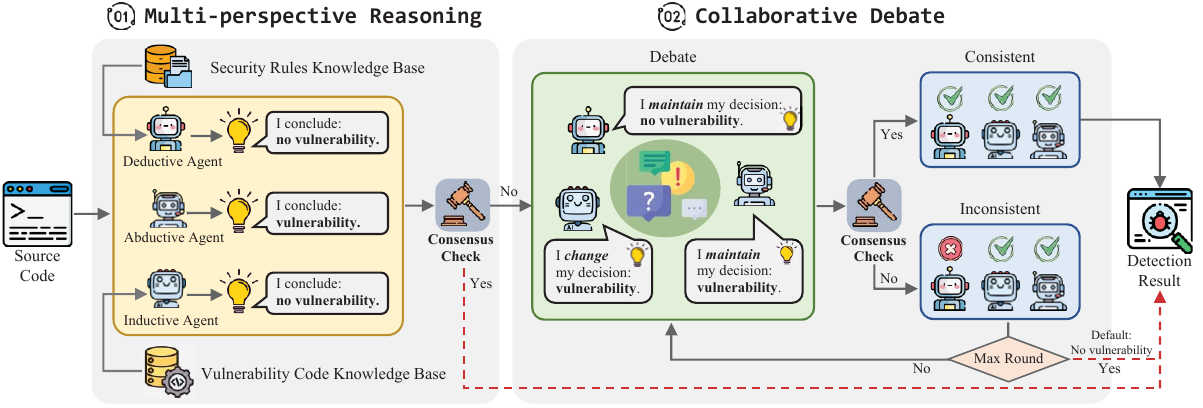}
\caption{The workflow of \toolname. }
\label{fig:framework}
\end{figure*}

\toolname leverages an LLM-based multi-agent system to achieve accurate vulnerability detection through cognitive synergy. 
The core idea is to simulate a team of security experts with diverse reasoning skills who collaborate to reach a reliable consensus. As illustrated in Fig.~\ref{fig:framework}, \toolname operates in two sequential stages:
\begin{itemize}[leftmargin=*]
    \item \textbf{Multi-perspective Reasoning:} Three specialist agents, each embodying a distinct reasoning paradigm: deductive, inductive, or abductive, independently analyze the input source code. This parallel analysis ensures a broad and diverse set of initial findings, maximizing the chances of uncovering different types of vulnerabilities.
    \item \textbf{Collaborative Debate:} If the agents' initial judgments conflict, they enter a multi-round debate. In this stage, they exchange, challenge, and defend their reasoning, iteratively refining their positions to resolve discrepancies and synthesize their unique insights.
\end{itemize}

This two-stage architecture is a direct response to our findings in Section~\ref{sec:motivation}. Relying on a single reasoning paradigm results in critical blind spots, while simplistic aggregation methods like majority voting are insufficient for resolving nuanced disagreements. By contrast, \toolname's debate mechanism facilitates a deeper integration of perspectives, leading to more accurate and robust vulnerability detection.

\subsection{Multi-perspective Reasoning}
The Multi-perspective Reasoning stage lays the groundwork for \toolname's collaborative analysis. Its primary objective is to generate a diverse set of independent assessments of a given source code snippet $X$. Each of the three agents, $\mathcal{A}_i$ (where $i \in \{D, I, A\}$ for Deductive, Inductive, and Abductive), is designed to operate in isolation, preventing premature convergence and ensuring a wide range of potential issues are considered.

To precisely define their functions, we formally model each agent's operation. An agent $\mathcal{A}_i$ acts as a function $f_i$ that takes the source code $X$ and a paradigm-specific knowledge source $K_i$ as input. It produces an output tuple $O_i = (v_i, e_i)$, where $v_i \in \{0, 1\}$ is the binary vulnerability judgment (1 for vulnerable, 0 for benign) and $e_i$ is a natural language explanation detailing its reasoning. This process is defined as $O_i = f_i(X, K_i)$. This formalization clarifies the distinct inputs and consistent output structure for each agent before we delve into their unique internal processes. 
Detailed information regarding the specific prompts and construction process for all agents can be found in the Data Availability section~\ref{sec:DataAvailability}, which provides the online repository.

\subsubsection{Deductive Agent}
The Deductive Agent emulates a security auditor who applies general principles to specific instances. It operates top-down, using established secure coding standards to identify violations. The rationale is that many vulnerabilities, such as integer overflows or improper resource handling, are direct violations of well-defined rules.

\textbf{Implementation.} The agent's analysis is powered by a Retrieval-Augmented Generation (RAG) mechanism designed to ground its reasoning in authoritative security knowledge. Its knowledge base, $K_D$, is derived from the SEI CERT C Coding Standard.~\footnote{\href{https://wiki.sei.cmu.edu/confluence/display/c/SEI\%2BCERT\%2BC\%2BCoding\%2BStandard}{https://wiki.sei.cmu.edu}} We structure it as a collection of pairs, $K_D = \{(\text{desc}_k, \text{rule}_k)\}_{k=1}^{M}$, where each $\text{desc}_k$ is the natural language description of a potentially flawed behavior (e.g., "Accessing freed memory corrupts heap data structures."), and $\text{rule}_k$ is the corresponding formal coding rule (e.g., "MEM30-C. Do not access freed memory").

During an offline indexing phase, we encode all natural language descriptions $\text{desc}_k$ into high-dimensional vectors using the Jina-V3 embedding model~\cite{sturua2024jina}, denoted as $\mathcal{E}(\cdot)$, creating an efficient retrieval index. When a source code snippet $X$ is provided for analysis, it is also embedded into a vector $\mathcal{E}(X)$. The system then computes a similarity score $s_k = \text{cos\_sim}(\mathcal{E}(X), \mathcal{E}(\text{desc}_k))$ for every description in the knowledge base. The indices of the top-5 descriptions with the highest scores, denoted by the set $\mathcal{J}^*$, are identified.


The corresponding set of rules, $\{ \text{rule}_k \mid k \in \mathcal{J}^* \}$, is then retrieved from the knowledge base. These rules are integrated into the final prompt, $P_D$, along with the input code $X$. The prompt's structure, guided by our template, instructs the Deductive Agent to act as a meticulous security auditor. It must systematically check if the code $X$ violates any of the provided high-relevance security rules and explain its findings. Finally, this context-rich prompt $P_D$ is processed by the agent's underlying LLM to produce the judgment and explanation $O_D = (v_D, e_D)$, ensuring the analysis is firmly grounded in established secure coding standards.

\subsubsection{Inductive Agent}
The Inductive Agent mirrors a data-driven security practitioner who generalizes from historical data to identify risks. It operates via a bottom-up, pattern-matching approach. The core intuition is that many software vulnerabilities are not entirely novel but are recurrences of previously discovered flawed patterns. By retrieving a known, analogous vulnerability instance, the agent can infer the risk in the new code snippet, much like an expert recalling a past incident to diagnose a present problem.

\textbf{Implementation.} The agent's reasoning is grounded in a knowledge base of real-world security issues. Following prior research~\cite{lin2025give}, we use the \textbf{ReposVul} dataset~\cite{wang2024reposvul}. This provides a knowledge base $K_I = \{(c_{vuln}^{(j)}, c_{fix}^{(j)})\}_{j=1}^{N}$ containing $N=6,956$ function-level vulnerability-fix pairs. Crucially, to ensure a fair evaluation and prevent data leakage, we meticulously filter this dataset, removing any examples that also appear in our downstream evaluation benchmarks.

The core of the agent's implementation is an RAG pipeline that retrieves the most relevant historical example. During an offline phase, every vulnerable function $c_{vuln}^{(j)} \in K_I$ is encoded into a high-dimensional vector using the Jina-V3 embedding model, which we denote as a function $\mathcal{E}(\cdot)$. When a new source code snippet $X$ is presented for analysis, it is also embedded into a vector $\mathcal{E}(X)$. The system then identifies the index $j^*$ of the most similar historical case by solving:
$j^* = \arg\max_{j} \text{cos\_sim}(\mathcal{E}(X), \mathcal{E}(c_{vuln}^{(j)}))$.
The corresponding pair, $(c_{vuln}^{(j^*)}, c_{fix}^{(j^*)})$, is then retrieved from the knowledge base to serve as a potent in-context example.
The final prompt, $P_I$, is then carefully constructed. It incorporates the retrieved pair $(c_{vuln}^{(j^*)}, c_{fix}^{(j^*)})$ as a concrete example alongside the input code $X$ that requires analysis. This entire context is framed by a set of predefined instructions from our prompt template, which guides the agent to first analyze the historical example to understand the vulnerability and its fix, and second, to determine if the new code snippet exhibits a similar flawed pattern. This prompt $P_I$ is then processed by the agent's underlying LLM to generate the final judgment and explanation $O_I = (v_I, e_I)$.

\subsubsection{Abductive Agent}
The Abductive Agent is designed to emulate the creative and exploratory mindset of a penetration tester. It reasons backward, starting from a hypothesized adverse outcome (an effect) and working to find the most plausible vulnerability in the code (the cause). This hypothesis-driven approach is uniquely suited for uncovering complex, context-dependent flaws such as race conditions, intricate logic bugs, or novel attack vectors that are not captured by predefined rules or historical patterns.

\textbf{Implementation.} The Abductive agent is centered on a sophisticated multi-step reasoning prompt that guides the LLM through a structured, internal thought process. Unlike its counterparts, $\mathcal{A}_A$ does not utilize an external knowledge base via RAG. Instead, it leverages the LLM's intrinsic, pre-trained knowledge of software, security, and attack patterns as its internal knowledge source, $K_A$. This allows it to generate novel hypotheses beyond the confines of any specific dataset.

The agent's prompt, $P_A$, instructs it to adopt the persona of an attacker and perform a structured CoT analysis on the input code $X$. The process is as follows: first, the agent must identify potentially risky code segments and generate a set of plausible attack hypotheses, $H = \{h_1, \dots, h_n\}$, regarding how the code might fail or be exploited. For each leading hypothesis $h \in H$, it is then required to construct a detailed reasoning trace, $T(h)$, outlining a potential attack path and explaining how specific lines of code contribute to the vulnerability. Finally, the agent must perform a crucial self-critique step, where it re-evaluates its own reasoning trace to assess its logical soundness and the real-world plausibility of the exploit. 

The final explanation, $e_A$, is a coherent synthesis of this entire internal monologue that goes from initial hypothesis to the validated conclusion. This structured, self-correcting reasoning process, encapsulated within the prompt $P_A$, enables the agent to form and validate complex arguments, ultimately leading to its final judgment $O_A = (v_A, e_A)$.

\begin{table}
\centering
\caption{The PairAcc (\%) of different LLMs on the PairVul dataset using three reasoning paradigms.}
\label{Tab:motivation-empirical}
\resizebox{0.8\textwidth}{!}{
\begin{tabular}{c|ccc|cccc} 
\toprule
\multirow{2}{*}{Paradigm} & \multicolumn{3}{c|}{Open Source}                 & \multicolumn{4}{c}{Commercial}                  \\ 
\cline{2-8}
                          & Llama4: 17b~   & Phi4: 14b~     & Devstral: 24b~ & GPT-4o~ & DeepSeek~      & Qwen3~ & Gemini 2.5  \\ 
\hline
Deductive                 & 15.17          & \textbf{21.89} & 18.41          & 18.41   & 19.15          & 19.70  & 14.43       \\
Inductive                 & \textbf{20.90} & 18.41          & 12.94          & 12.19   & 18.41          & 16.17  & 10.95       \\
Abductive                 & 17.41          & 17.41          & 13.43          & 11.19   & \textbf{22.14} & 15.42  & 6.72        \\
\bottomrule
\end{tabular}
}
\end{table}

\subsubsection{Model Specialization}
Recognizing that a one-size-fits-all approach is suboptimal, and that LLM architecture and training data lead to varying aptitudes for different reasoning tasks~\cite{yin2024multitask, steenhoek2024err}, we conducted an empirical study to assign the best-performing model to each reasoning paradigm. Using the PairVul dataset~\cite{Vul-rag}, we evaluated model performance with PairAcc. Our experimental results under this metric, summarized in Table~\ref{Tab:motivation-empirical}, revealed clear specializations. Consequently, we configured our agents to leverage these model-specific strengths: the Deductive Agent is powered by Phi4 (21.89\% PairAcc), the Inductive Agent employs Llama4 (20.90\%), and the Abductive Agent leverages DeepSeek (22.14\%). This data-driven assignment ensures that each agent operates at its peak empirical performance, thereby maximizing the quality and diversity of the initial judgments before the collaborative debate begins.

\subsubsection{Phase Conclusion and Transition}
The Multi-perspective Reasoning phase concludes once all three specialized agents have produced their independent outputs. These results are aggregated into a collective set, $\mathcal{O}^0 = \{O_D^0, O_I^0, O_A^0\}$. The framework then determines the next step based on a transition variable, $\mathcal{T}^0$, which checks for consensus:
\[
\mathcal{T}^0 = \begin{cases} 
\text{exit}, & \text{if } v_D^0 = v_I^0 = v_A^0 \\
\text{debate}, & \text{otherwise}
\end{cases}
\]
If all agents are in agreement ($\mathcal{T}^0 = \text{exit}$), their consistent judgment $v_f = v_D^0$ is accepted as the final result, and their individual explanations $\{e_D^0, e_I^0, e_A^0\}$ are synthesized into a single, coherent report. However, if there is any disagreement ($\mathcal{T}^0 = \text{debate}$), the framework advances to the Collaborative Debate phase to resolve the conflict.

\subsection{Collaborative Debate}
While the Multi-perspective Reasoning phase generates a crucial diversity of initial opinions, these independent analyses are often insufficient to resolve complex cases where paradigms conflict. Relying on simplistic aggregation methods, such as majority voting, risks discarding the nuanced insights of a dissenting agent. This phase realizes our key insight that cognitive synergy is achieved through iterative interaction. We introduce the Collaborative Debate, a structured process designed to simulate the scientific principle that truth emerges from critical discourse~\cite{liu2025truth}. By compelling agents to confront, justify, and synthesize differing viewpoints, the debate forces a re-examination of evidence that may have been initially overlooked or misinterpreted. This iterative refinement is the engine that drives the agents toward a unified consensus that is more robust and reliable than any of their isolated, initial judgments.

The debate commences if the initial judgments diverge and proceeds in rounds, $t = 1, 2, \dots, T_{\max}$. A core design principle is that agents deliberate and update their views in parallel within each round. This ``simultaneous speaking'' model prevents sequential bias, where the order of speakers could unduly influence the outcome, and ensures each reasoning paradigm is given equal weight~\cite{estornell2024multi}. In any given round $t$, each agent $\mathcal{A}_i$ receives a comprehensive context for its deliberation. This context includes the source code $X$, its specific knowledge $K_i$, its own reasoning history $\{O_i^0, \dots, O_i^{t-1}\}$, and the most recent outputs from its peers, $\{O_j^{t-1} \mid j \neq i\}$.

Within each round, every agent performs a parallel deliberation process. It first critically audits the reasoning chains ($e_j^{t-1}$) of its peers, especially those with conflicting judgments ($v_j^{t-1} \neq v_i^{t-1}$). It then re-evaluates the evidence through the lens of its own paradigm. If a peer's argument is persuasive and reveals a flaw in its own prior logic, the agent revises its stance. Conversely, if it finds the peer's reasoning unsound, it formulates a rebuttal, reinforcing its original position and attempting to persuade others. This deliberation process, which we denote as the function $g_i$, generates an updated output tuple for the agent: $O_i^t = (v_i^t, e_i^t) = g_i(X, K_i, \{O_i^0, \dots, O_i^{t-1}\}, \{O_j^{t-1} \mid j \neq i\})$.

After each round of deliberation, the framework aggregates the new outputs into a set $\mathcal{O}^t = \{O_D^t, O_I^t, O_A^t\}$ and checks for consensus by evaluating the transition state $\mathcal{T}^t$:
\[
\mathcal{T}^t = \begin{cases} 
\text{exit}, & \text{if } v_D^t = v_I^t = v_A^t \\
\text{debate}, & \text{otherwise}
\end{cases}
\]
If consensus is reached ($\mathcal{T}^t = \text{exit}$), the debate terminates. The unanimous decision becomes the final judgment, $v_f$, and a final explanation, $e_f$, is synthesized from the agents' convergent reasoning chains. If no consensus is reached after $T_{\max}$ rounds, the framework defaults to a non-vulnerable judgment ($v_f = 0$). This is a deliberate design choice, motivated by empirical findings that LLMs are prone to false positives in vulnerability detection~\cite{zibaeirad2024vulnllmeval, VulTrial} and aligns with the security principle of minimizing alert fatigue.

\section{Experiment Design}
\label{sec:setup}
\subsection{Research Questions}
To evaluate the effectiveness of our approach, we address the following research questions:
\begin{itemize}
    \item \textbf{RQ1:} How does \toolname compare to state-of-the-art vulnerability detection approaches?
    \item \textbf{RQ2:} How is the generalizability of \toolname in software vulnerability detection?
    \item \textbf{RQ3:} What are the contributions of the major components of \toolname?
\end{itemize}

\subsection{Dataset}
We evaluate \toolname on the PrimeVul~\cite{PrimeVul} dataset, a high-quality function-level benchmark for vulnerability detection that addresses label inaccuracies through combined automated labeling and manual verification, coupled with rigorous de-duplication. PrimeVul features diverse vulnerabilities across 140 CWE categories, sourced from large-scale open-source projects, with a test set of 435 vulnerable/fixed code pairs. To further validate generalization (RQ2), we utilize the JITVUL~\cite{yildiz2025benchmarking} dataset, a repository-level benchmark derived from real-world projects, providing interprocedural context via caller-callee relationships. JITVUL includes 1758 paired samples, comprising 879 vulnerable and 879 benign versions, spanning 91 CWEs from 879 CVEs.

\subsubsection{Baselines}
We select baselines from two major categories, covering representative state-of-the-art techniques in vulnerability detection: learning-based methods and LLM-based methods. The learning-based baselines include CausalVul~\cite{rahman2024towards}, DeepDFA~\cite{steenhoek2024dataflow}, VulChecker~\cite{mirsky2023vulchecker}, FVD-DPM~\cite{shao2024fvd}, Coca~\cite{coca}, VulSim~\cite{shimmi2024vulsim}, LineVul~\cite{linevul}, and MoEVD~\cite{MoEVD}. The LLM-based baselines include SAVul~\cite{SAVul}, Vul-RAG~\cite{Vul-rag}, VulnSage~\cite{VulnSage}, GPTLens~\cite{GPTLens}, and VulTrial~\cite{VulTrial}.
To ensure a fair comparison, all baseline methods were evaluated under the same experimental environment and were re-executed using their officially provided publicly available implementations.

\subsection{Evaluation Metrics}
To comprehensively evaluate the performance of \toolname in vulnerability detection, we employ a suite of metrics, with a primary focus on PairAcc, as it is currently the most prominent and representative evaluation metric in the vulnerability detection domain~\cite{yildiz2025benchmarking,PrimeVul,VulTrial}. 
PairAcc imposes a dual requirement that necessitates the correct classification of both a vulnerable function and its fixed version. This metric essentially serves as a strict standard that balances the need for high Recall with the necessity of a low False Positive Rate. In practical scenarios, suppressing false positives is crucial to maintaining developer trust and reducing alert fatigue. By requiring the model to distinguish between the vulnerable and fixed code, PairAcc effectively penalizes models that tend to aggressively predict vulnerabilities, ensuring that the detector possesses the discriminative capability required to minimize false alarms on benign code.
For each (vulnerable, fixed) pair, the model must predict the former as ``1'' and the latter as ``0''. Formally, PairAcc is defined as:
\begin{equation}
\text{PairAcc} = \frac{|\{(v, f) \in \mathcal{P} \mid \hat{y}_v = 1 \land \hat{y}_f = 0\}|}{|\mathcal{P}|},
\end{equation}
where \(\mathcal{P}\) represents the set of all (vulnerable, fixed) code pairs, \((v, f)\) denotes a single pair with \(v\) as the vulnerable code and \(f\) as the fixed code, \(\hat{y}_v\) and \(\hat{y}_f\) are the predictions for the vulnerable and fixed code snippets, respectively, and \(|\mathcal{P}|\) is the total number of pairs. In addition to PairAcc, we report widely-used classification metrics: \textbf{Accuracy}: Accuracy measures the overall correctness of predictions. \textbf{Precision}: Precision measures the proportion of true vulnerabilities among all predicted vulnerable instances. \textbf{Recall}: Recall measures the proportion of actual vulnerabilities correctly identified.  \textbf{F1-score}: F1-score measures the harmonic mean of precision and recall. \textbf{False Positive Rate (FPR)}: FPR measures the proportion of non-vulnerable code incorrectly predicted as vulnerable.

\subsection{Implementation Details}
For open-source LLMs, we implement \toolname on a server equipped with four NVIDIA RTX 4090 GPUs. 
We download the model weights of Llama-4~\footnote{\url{https://huggingface.co/meta-llama/Llama-4-Scout-17B-16E-Instruct}}, Phi-4~\footnote{\url{https://huggingface.co/microsoft/phi-4}}, and Devstral~\footnote{\url{https://huggingface.co/mistralai/Devstral-Small-2505}} from the HuggingFace and deploy them locally.
For commercial LLMs, we access the models directly through their official APIs. Specifically, we employ the following model versions: gpt-4o-2024-05-13, qwen-max, gemini-2.5-flash, and deepseek-v3-0324. Regarding the generation configuration, we set the \texttt{temperature} to 0 to minimize randomness and ensure the reproducibility of our results. For all other hyperparameters, we adopt the default settings provided by the openai library~\footnote{\url{https://github.com/openai/openai-python}}: \texttt{top\_p} is set to 1.0, \texttt{top\_k} is set to 50, and \texttt{repetition\_penalty} is set to 1.0.
For the parameter settings of \toolname, we explored the maximum number of debate rounds \( T_\text{max} \) within a selection space of 1 to 5. Based on subsequent ablation experiments, we ultimately set \( T_\text{max} \) to 2.

\section{Experiment Results}
\subsection{RQ1: Compare with State-of-the-Art Approaches}
To answer RQ1, we compare ReasonVul with a comprehensive set of state-of-the-art baseline methods, including both learning-based and LLM-based approaches. For learning-based methods, we train each model on the PrimeVul training set using the official hyperparameter configurations provided by the paper. For models with limited context window capacity, inputs that exceed the maximum supported context length are truncated following the constraints of the original implementations. For example, CausalVul, LineVul, and MoEVD are implemented based on CodeBERT, which supports a maximum input length of 512 tokens. Note that we adapted the line-level tool (i.e., VulChecker) to the function-level detection task to ensure consistency of metrics across methods. The trained models are then evaluated on the PrimeVul test set using the learned weights. For LLM-based methods, we directly apply their officially released prompts and pipelines to the PrimeVul test set without additional fine-tuning. Finally, we report PairAcc, accuracy, precision, recall, F1-score, and FPR for all methods.

\begin{table}
\centering
\caption{Comparison with state-of-the-art vulnerability detection techniques on the PrimeVul Dataset}
\label{tab:RQ1-SOTA}
\resizebox{0.99\linewidth}{!}{
\begin{tabular}{l|c|cccccc} 
\toprule
Type                            & Method   & PairAcc ($\uparrow$)       & Accuracy ($\uparrow$)      & Precision ($\uparrow$)     & Recall ($\uparrow$)   & F1-score ($\uparrow$) & FPR ($\downarrow$)        \\ 
\hline
\multirow{8}{*}{Learning-based} & CausalVul  & 5.74    & 49.43    & 49.44     & 51.03  & 50.23    & 52.18  \\
                                & DeepDFA    & 4.36    & 49.54    & 49.51     & 46.21  & 47.80    & 47.12  \\
                                & VulChecker & 3.67    & 48.28    & 47.81     & 37.70  & 42.16    & 41.15  \\
                                & FVD-DPM    & 5.52    & 52.30    & 53.28     & 44.83  & 48.69    & 39.31  \\
                                & Coca       & 4.14    & 50.92    & 50.91     & 51.95  & 51.42    & 50.11  \\
                                & VulSim     & 8.51    & 49.20    & 49.28     & 54.71  & 51.85    & 56.32  \\
                                & LineVul    & 4.36    & 52.07    & 52.31     & 46.90  & 49.45    & 42.76  \\
                                & MoEVD      & 8.97    & 53.67    & 54.12     & 48.27  & 51.03    & 40.92  \\  
\hline
\multirow{5}{*}{LLM-based}      & SAVul      & 11.03   & 53.91    & 55.28     & 40.92  & 47.28    & 37.93  \\
                                & Vul-RAG    & 10.57   & 51.95    & 52.00     & 51.03  & 51.51    & 47.13  \\
                                & VulnSage   & 8.28    & 48.97    & 48.31     & 29.66  & 36.75    & 31.72  \\
                                & GPTLens    & 13.56   & 51.73    & 51.20     & 78.16  & 61.87    & 74.71  \\
                                & VulTrial   & 22.07   & 55.17    & 54.95     & 57.47  & 56.18    & 47.13  \\ 
\hline
LLM-based                       & \toolname   & 40.00   & 69.77    & 66.48     & 79.77  & 72.52    & 40.23  \\
\bottomrule
\end{tabular}
}

\end{table}

Table~\ref{tab:RQ1-SOTA} summarizes the performance of \toolname and the baseline methods on the PrimeVul dataset. Overall, our experimental results demonstrate that \toolname achieves superior performance across all evaluation metrics. Specifically, \toolname attains a PairAcc of 40.00\%, indicating its exceptional ability to accurately distinguish between vulnerable and fixed code pairs, capturing subtle semantic differences that characterize real-world vulnerabilities. This result represents substantial improvements over the best-performing baseline (VulTrial), with gains of 81.24\% in PairAcc. This highlights the effectiveness of our multi-agent debate framework in leveraging diverse reasoning paradigms for more accurate and reliable vulnerability detection.

Learning-based methods exhibit moderate performance, with PairAcc values ranging from 3.67\% to 8.97\% and F1-scores around 42\%-52\%. These approaches, reliant on supervised learning from labeled datasets, are limited by their focus on syntactic and structural patterns, which hinders their ability to distinguish subtle semantic differences between vulnerable and fixed code pairs. Their low PairAcc values reflect this limitation, as they struggle to generalize to unseen vulnerabilities or capture the nuanced characteristics required for accurate pairwise detection.
The LLM-based methods generally demonstrate better results than the learning-based methods, due to their powerful code understanding capabilities, with PairAcc ranging from 8.28\%-10.57\%. Specifically, VulnSage, which leverages an abductive reasoning approach through a self-reflection and validation mechanism, tends to output benign results, resulting in a lower recall rate. Vul-RAG, employing an inductive reasoning strategy with RAG, improves performance by increasing the recall rate to 51.03\% and achieving an F1-score of 51.51\%. However, these methods are still limited by the inherent reasoning capabilities of a single LLM, and their performance remains constrained when dealing with complex code semantics. The GPTLens and VulTrial outperform the previous approaches, with PairAcc values of 13.56\% and 22.07\%, respectively. 
Notably, while GPTLens achieves a high recall rate of 78.16\%, it suffers from a significantly high FPR of 74.71\%, indicating a severe tendency for over-detection. In contrast, VulTrial introduces roles for a judge and a jury to simulate a courtroom debate mechanism, improving the balance of performance. However, \toolname achieves a much lower FPR of 40.23\% while maintaining high recall, demonstrating superior capability in reducing false alarms compared to adversarial baselines.
Therefore, \toolname demonstrates superior vulnerability detection capability compared to the baseline methods.

\begin{table}[!htbp]
\centering
\caption{Accuracy Comparison on Top-10 Most Frequent CWEs in PrimeVul.}
\label{tab:cwe_acc}
\resizebox{0.9\linewidth}{!}{
\begin{tabular}{llcccc}
\toprule
CWE ID & Vulnerability Type & Count & GPTLens & VulTrial & \toolname \\
\midrule
CWE-787 & Out-of-bounds Write & 144 & 53.47 & 54.86 & 66.67 \\
CWE-125 & Out-of-bounds Read & 94 & 53.19 & 54.26 & 75.53 \\
CWE-703 & Improper Check/Handling of Exceptions & 94 & 56.38 & 51.06 & 72.04 \\
CWE-476 & NULL Pointer Dereference & 78 & 48.72 & 53.85 & 75.64 \\
CWE-416 & Use After Free & 58 & 41.38 & 53.45 & 68.97 \\
CWE-200 & Exposure of Sensitive Information & 32 & 53.12 & 62.50 & 59.38 \\
CWE-20 & Improper Input Validation & 28 & 53.57 & 50.00 & 78.57 \\
CWE-369 & Divide By Zero & 28 & 57.14 & 46.43 & 67.86 \\
CWE-119 & Improper Restriction of Operations & 28 & 53.57 & 57.14 & 71.43 \\
CWE-617 & Reachable Assertion & 24 & 62.50 & 54.17 & 70.83 \\
\midrule
\textbf{Average} &   &   & 53.30 & 53.77 & 70.69 \\
\bottomrule
\end{tabular}
}
\end{table}

\subsubsection{Performance across different vulnerability types}
To assess the efficacy of ReasonVul in addressing real-world security threats, we conducted a thorough investigation into the Top-10 most frequent CWEs in the PrimeVul dataset. These vulnerabilities represent the most common occurrences in our dataset and pose significant risks due to their inherent ease of discovery and exploitation. We compare the accuracy of ReasonVul against two LLM-based baselines, GPTLens and VulTrial.

As shown in Table~\ref{tab:cwe_acc}, \toolname consistently outperforms the baselines across nearly all frequent CWE categories, achieving an average accuracy of 70.69\%, which is substantially higher than GPTLens at 53.30\% and VulTrial at 53.77\%. This advantage indicates that \toolname is not merely improving the aggregate performance reported in Table~\ref{tab:RQ1-SOTA}, but also provides more stable detection capability when the vulnerabilities are grouped by their underlying weakness types. 
Overall, these results verify that the cognitive synergy of deductive, inductive, and abductive reasoning effectively addresses diverse vulnerability characteristics, offering superior robustness over role-playing approaches.

\begin{tcolorbox}
\textbf{Answer to RQ1:} \toolname significantly outperforms all baseline methods in vulnerability detection, achieving the highest scores across all metrics. Compared to previous LLM-based multi-agent approaches, PairAcc improves by 81.24\%-194.99\%.
\end{tcolorbox}

\subsection{RQ2: Generalizability of \toolname}
We have demonstrated that \toolname achieves impressive performance, successfully detecting vulnerabilities from the widely used PrimeVul dataset. To assess the generalizability of \toolname, we extend our evaluation from the function-level PrimeVul dataset to the more challenging repository-level JITVUL dataset. While PrimeVul provides clean, function-level code snippets, JITVUL introduces a higher degree of complexity by providing repository-level context, including interprocedural dependencies like caller-callee relationships. This setup more closely mirrors real-world code auditing scenarios where vulnerabilities often arise from interactions between different parts of a codebase (i.e., interprocedural vulnerabilities). In this RQ, we investigate whether \toolname can effectively detect interprocedural vulnerabilities given a suspicious function and its immediate calling context (i.e., its caller and callee functions). 

To capture interprocedural dependencies, we employ CFlow~\cite{GNU} to statically analyze the source code and extract the caller and callee functions associated with each target function. Following the setting of prior works~\cite{wen2024vuleval,yildiz2025benchmarking}, when multiple callers or callees exist, we retrieve the top-5 most similar callers and the top-5 most similar callees, and serialize them into a unified input sequence in the format of \textit{[Caller Context]} + \textit{[Target Function]} + \textit{[Callee Context]}. The contextual information for each caller and callee includes both its function signature and function body. This structured input ensures that \toolname receives the necessary interprocedural contextual information to reason across procedure boundaries.
To ensure the fairness of the experiment, we follow the same experimental setup as in RQ1, using identical baselines. 
In particular, for learning-based methods, we directly reuse the model weights trained on the PrimeVul training set when evaluating on the JITVUL dataset.

\begin{table}
\centering
\caption{Performance comparison of repository-level vulnerability detection on the JITVUL dataset}
\label{tab:RQ2-general}
\resizebox{0.99\linewidth}{!}{
\begin{tabular}{l|c|cccccc} 
\toprule
Type & Method   & PairAcc ($\uparrow$)       & Accuracy ($\uparrow$)      & Precision ($\uparrow$)     & Recall ($\uparrow$)   & F1-score ($\uparrow$) & FPR ($\downarrow$)        \\ 
\hline
\multirow{8}{*}{Learning-based} & CausalVul  & 3.07    & 51.54    & 51.50     & 52.90  & 52.19    & 49.83  \\
                                & DeepDFA    & 3.98    & 51.99    & 52.25     & 46.19  & 49.03    & 42.21  \\
                                & VulChecker & 3.19    & 51.59    & 51.83     & 45.05  & 48.20    & 41.87  \\
                                & FVD-DPM    & 4.32    & 52.16    & 52.11     & 53.36  & 52.73    & 49.03  \\
                                & Coca       & 2.05    & 50.85    & 50.49     & 87.71  & 64.09    & 86.01  \\
                                & VulSim     & 6.94    & 53.47    & 53.17     & 58.25  & 55.59    & 51.31  \\
                                & LineVul    & 5.12    & 51.65    & 51.75     & 48.69  & 50.18    & 45.39  \\
                                & MoEVD      & 5.80    & 52.28    & 51.85     & 63.82  & 57.22    & 59.27  \\
\hline
\multirow{5}{*}{LLM-based}      & SAVul      & 8.99    & 49.77    & 49.84     & 68.83  & 57.81    & 69.28  \\
                                & Vul-RAG    & 12.97   & 54.32    & 53.51     & 65.87  & 59.05    & 57.22  \\
                                & VulnSage   & 9.22    & 50.17    & 50.27     & 32.08  & 39.17    & 31.74  \\
                                & GPTLens    & 10.69   & 52.90    & 51.68     & 89.42  & 65.50    & 82.13  \\
                                & VulTrial   & 20.93   & 56.83    & 56.93     & 56.09  & 56.50    & 42.43  \\
\hline
LLM-based                       & \toolname  & 28.67   & 63.54    & 61.85     & 70.65  & 65.96    & 43.57     \\
\bottomrule
\end{tabular}
}

\end{table}

As shown in Table~\ref{tab:RQ2-general}, \toolname achieves the highest performance across all metrics, with a PairAcc of 28.67\% and an F1-score of 65.96\%. Although all methods experience a performance drop compared to PrimeVul due to JITVUL’s increased complexity, \toolname maintains a significant advantage over prior approaches. 
Specifically, compared with VulSim, which achieves the best performance among learning-based methods, \toolname attains an absolute improvement of 21.73\% in PairAcc and 10.37\% in F1-score. Compared with VulTrial, which achieves the best performance among LLM-based methods, \toolname achieves an absolute improvement of 7.74\% in PairAcc and 9.46\% in F1-score.
Notably, learning-based methods exhibit a noticeable performance degradation compared to their evaluation results in RQ1. Specifically, token-based methods constrained by context length limits (e.g., 512 tokens), truncate critical inter-procedural information~\cite{jiang2024stagedvulbert}, while graph-based methods fail when handling large-scale code graphs~\cite{wen2023vulnerability}.

Among LLM-based methods, we observe that the F1-score improvement of \toolname over GPTLens is only 0.46\%. However, a comprehensive analysis across all evaluation metrics reveals significant differences in detection capabilities. GPTLens primarily achieves high recall through aggressive predictions, which results in an extremely high FPR of 82.13\% and leads to an inflated performance under aggregate metrics. In contrast, \toolname attains a higher F1-score while reducing the FPR by 38.56\%. These results indicate that, compared to GPTLens, \toolname exhibits a stronger capability to detect vulnerabilities that require reasoning over interprocedural context.

\begin{tcolorbox}
\textbf{Answer to RQ2:} \toolname exhibits strong generalizability on the JITVUL dataset, significantly outperforming baselines in repository-level vulnerability detection, with improvements of 65.91\% in PairAcc over the best baseline VulTrial.
\end{tcolorbox}

\subsection{Ablation Study}
To dissect the contributions of \toolname's core components, we conduct ablation studies on the PrimeVul dataset, focusing on two key aspects: the impact of individual reasoning agents and the influence of debate rounds. 
\subsubsection{The impact of each reasoning thought}
We first evaluate the contribution of each reasoning agent by comparing the full \toolname framework against configurations that either exclude one agent or use a single agent in isolation. Table~\ref{tab:RQ3-1-ablation} reports the performance metrics of different variants on the PrimeVul dataset. The experimental results show that the full \toolname configuration, which incorporates all three agents, achieves the highest performance across all metrics, underscoring the critical role of cognitive diversity in vulnerability detection. Removing any single agent results in a consistent performance decline, demonstrating that each agent contributes uniquely to the framework’s effectiveness. Specifically, the impact of disabling the deductive agent is the smallest, resulting in a PairAcc of 35.00\% and an F1-score of 66.67\%, representing decreases of 5.00\% and 5.85\% compared to the full model. Going further, removing the inductive agent reduces PairAcc to 31.95\% and F1-score to 66.45\%, corresponding to drops of 8.05\% and 6.07\%. Finally, the most significant performance degradation occurs when the abductive agent is removed, with PairAcc falling to 24.60\% and F1-score to 59.78\%, decreases of 15.40\% and 12.74\%, respectively. This underscores the Abductive Agent's unique ability to identify complex, context-dependent vulnerabilities that other paradigms miss. Additionally, we observed that all single-agent configurations performed significantly worse than the full model, with PairAcc dropping by 12.91\%–19.08\% and F1-score decreasing by 10.38\%–12.55\%. This confirms that the integration of deductive, inductive, and abductive reasoning is essential for effectively capturing the diversity of vulnerabilities.


\begin{table}
\centering
\caption{The impact of individual reasoning agents}
\label{tab:RQ3-1-ablation}
\resizebox{0.75\linewidth}{!}{
\begin{tabular}{l|ccccc} 
\toprule
Configuration         & PairAcc               & Accuracy             & Precision            & Recall               & F1-score              \\ 
\hline
w/o Deductive Agent & 35.00   & 64.45    & 68.09     & 65.31  & 66.67     \\
w/o Inductive Agent & 31.95   & 65.17    & 64.10     & 68.97  & 66.45     \\
w/o Abductive Agent & 24.60   & 57.47    & 56.70     & 63.22  & 59.78     \\ 
\hline
Deductive Agent~Only~ & 25.28                 & 58.16                & 57.17                & 65.06                & 60.86                 \\
Inductive Agent~Only~ & 20.92                 & 55.06                & 67.36                & 67.36                & 59.97                 \\
Abductive Agent~Only~ & 27.81                 & 62.18                & 62.21                & 62.07                & 62.14                 \\ 
\hline
Full                  & 40.00                 & 69.77                & 66.48                & 79.77                & 72.52                 \\
\bottomrule
\end{tabular}
}

\end{table}
\subsubsection{The impact of debate}
\label{subsec:debate}
\begin{wrapfigure}[14]{r}{0.50\linewidth}
  \centering
  \includegraphics[width=\linewidth]{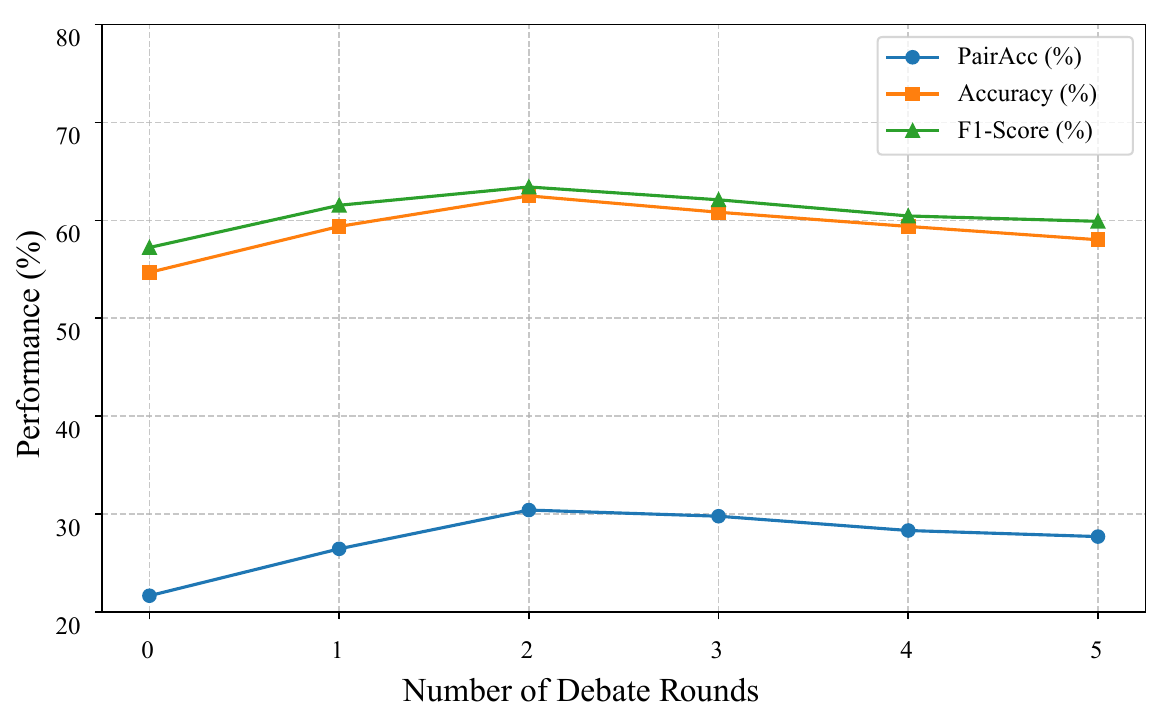}
  \caption{The impact of debate rounds}
  \label{fig:debate-rounds}
\end{wrapfigure}

We investigate the impact of varying the maximum number of debate rounds $T_{\max}$ from 0 to 5 on the PrimeVul validation set, where $T_{\max}=0$ corresponds to relying solely on the multi-perspective reasoning phase with majority voting to resolve conflicts. The results, illustrated in Fig.~\ref{fig:debate-rounds}, reveal performance trends for PairAcc, Accuracy, and F1-score across these settings. At 0 rounds, PairAcc is 21.67\%, Accuracy is 54.69\%, and F1-score is 57.23\%, highlighting the limitations of majority voting, which fails to exploit nuanced insights embedded in conflicting agent judgments.

Introducing debate substantially improves performance: with just 1 round, PairAcc increases to 26.46\%, Accuracy to 59.38\%, and F1-score to 61.54\%. The best results are achieved at 2 rounds, where PairAcc reaches 30.42\%, Accuracy 62.50\%, and F1-score 63.41\%, demonstrating that structured debate effectively promotes cognitive synergy through iterative refinement. 
Beyond 2 rounds, performance begins to decline. This decline is attributed to the excessively long context accumulated in multi-round dialogues, which can overwhelm the ability of agents to maintain coherent reasoning, leading to potential misinterpretations or conservative judgments~\cite{li2023loogle}.
These results validate our choice of $T_{\max}=2$, which optimizes detection accuracy while maintaining computational efficiency.

In addition, to validate our design choice of defaulting to a benign judgment when consensus is not reached after $T_{\max}$ rounds, we conducted experiments on the test set that showed an absolute improvement of 4.14\% in PairAcc compared to the majority voting strategy. These results indicate that adopting a conservative strategy after debate reduces false positives, which aligns with prior findings on the over-detection tendencies of LLMs~\cite{zibaeirad2024vulnllmeval}.

\begin{tcolorbox}
\textbf{Answer to RQ3:} Each reasoning agent contributes uniquely, with their combination yielding substantial gains. The debate mechanism is crucial, with 2 rounds optimizing performance by enabling effective cognitive synergy.
\end{tcolorbox}

\section{Discussion}
\subsection{Case Study}
\begin{figure*}[!t]
\centering
\includegraphics[width=0.99\textwidth]{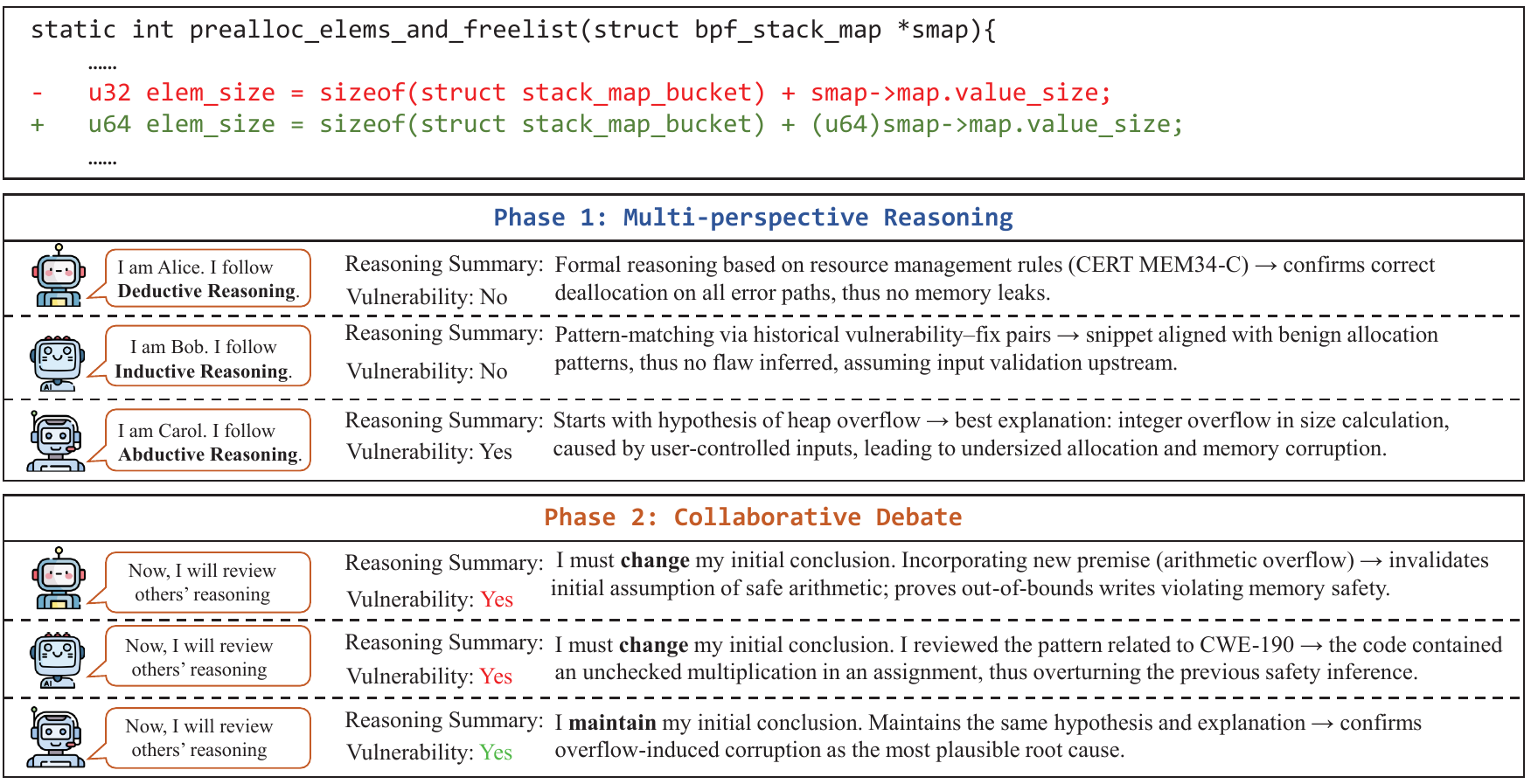}
\caption{Case study of the \toolname for CVE-2021-41864.}
\label{fig:debate-example}
\end{figure*}

To illustrate the critical role of the collaborative debate mechanism in \toolname, we present a case study analyzing a real-world integer overflow vulnerability from the Linux kernel~\footnote{\url{https://nvd.nist.gov/vuln/detail/CVE-2021-41864}}. This case is selected to demonstrate how integrating multiple reasoning paradigms, combined with structured debate, enables \toolname to detect subtle vulnerabilities that might be overlooked by any single reasoning method or by simple aggregation approaches such as majority voting. The condensed transcript of the agent interactions is depicted in Fig.~\ref{fig:debate-example}, with concise reasoning summaries generated by the LLM for clarity.

In the Multi-perspective Reasoning phase, each agent independently analyzes the code, leveraging its unique reasoning paradigm. The Deductive Agent, applying the formal security principle of resource management (CERT MEM34-C), confirms that the function's control flow correctly deallocates memory on all error paths, concluding that the code is compliant with this specific rule and shows no evidence of memory leaks. However, this focus on memory management overlooks the subtle arithmetic boundary violation causing the integer overflow. The Inductive Agent, relying on pattern-matching against historical vulnerability data from the ReposVul dataset, does not detect the issue, as the specific overflow pattern is not well-represented in its knowledge base. Only the Abductive Agent, through hypothesis-driven reasoning, identifies the vulnerability by postulating a failure scenario involving excessive user input that could trigger an integer overflow, leading to an out-of-bounds access. This divergence highlights a key insight: relying on a single reasoning paradigm is insufficient for detecting the complex and diverse vulnerabilities that arise in real-world scenarios.

The Collaborative Debate phase is pivotal in synthesizing these diverse perspectives and resolving the initial 2-to-1 conflict (Deductive and Inductive Agents voting "benign" versus Abductive Agent voting "vulnerable"). As depicted in Fig.~\ref{fig:debate-example}, a simple majority voting approach at this stage would have erroneously classified the code as benign, overlooking the vulnerability. Instead, \toolname's structured debate mechanism enables dynamic error correction. The Abductive Agent presents its hypothesis-driven evidence, articulating a plausible attack scenario that highlights the overflow risk. This prompts the Deductive Agent to re-evaluate the code beyond CERT MEM34-C, recognizing a previously overlooked violation in arithmetic safety principles. Similarly, when confronted with the abductive hypothesis, the inductive agent re-evaluates its pattern-matching approach, acknowledges that it previously overlooked this vulnerability pattern, and adjusts its stance to align with the new evidence. Through this iterative dialogue, all agents are ultimately able to correctly identify the vulnerability. To further demonstrate the effectiveness of collaborative debate, we compared its performance with majority voting in resolving conflicting judgments. Across 542 samples with initial conflicts in the PrimeVul dataset, majority voting successfully resolved only 49 cases, accounting for a mere 9.04\%. In contrast, the collaborative debate mechanism correctly resolved 389 cases, representing 71.77\%. This striking disparity underscores the capability of the debate mechanism to emulate the critical, multi-perspective reasoning of human expert panels, enabling agents to challenge and refine their analyses through evidence-based discourse.

This case study, coupled with the quantitative comparison, highlights two key strengths of \toolname. First, multi-perspective reasoning is essential, as the complementary nature of deductive, inductive, and abductive approaches ensures that diverse aspects of complex vulnerabilities are captured, which cannot be achieved by a single paradigm alone. Second, the collaborative debate mechanism turns initial differences in opinion into a chance to improve detection accuracy, facilitating a dynamic process of discovery and error correction that significantly outperforms simplistic aggregation methods like majority voting. Compared to single-paradigm methods or basic ensemble approaches, \toolname's integration of multi-perspective reasoning and structured debate substantially enhances the accuracy of vulnerability detection.

\subsection{Analysis of Collaborative Synergy}
To quantify the synergy generated by our framework, this section analyzes how the collaborative debate synthesizes the diverse outputs from the individual Deductive, Inductive, and Abductive agents. 
The analysis, based on the PrimeVul dataset evaluation, reveals that the debate mechanism is crucial for integrating complementary perspectives into a unified and more accurate judgment.

\begin{wrapfigure}[16]{r}{0.5\linewidth}
  \centering
  \includegraphics[width=\linewidth]{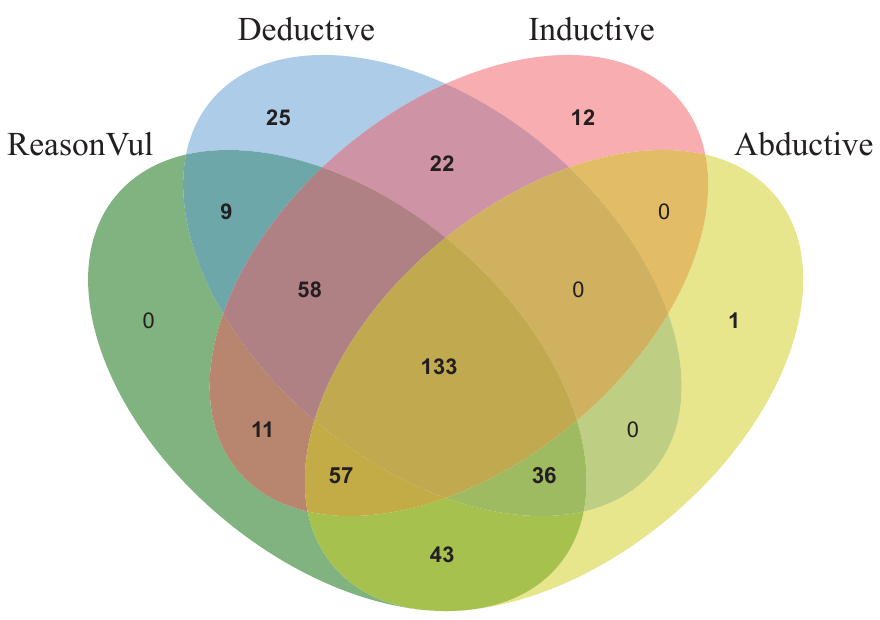}
  \caption{The overlaps of the vulnerabilities detected by different approaches.}
  \label{fig:venn-full-vs-agents}
\end{wrapfigure}

Fig.~\ref{fig:venn-full-vs-agents} illustrates the complementarity among the agents' initial findings. Before the debate, the deductive, abductive, and inductive agents independently detected 283, 293, and 270 vulnerabilities, respectively, with their intersection limited to just 133 vulnerabilities. This overlap confirms that each reasoning paradigm possesses unique strengths but is also hindered by notable blind spots. After the debate, however, \toolname successfully identifies a total of 347 vulnerabilities, of which 214 cases represent instances where the debate effectively resolved initial disagreements. This demonstrates that the debate mechanism effectively transforms fragmented and often conflicting initial analyses into a consistent and far more comprehensive final result.
\toolname outperforms the individual agents by detecting 64, 54, and 77 additional vulnerabilities compared to the deductive, abductive, and inductive agents, respectively. This enhanced coverage highlights the framework's capacity to leverage the collective intelligence of its agents, uncovering subtle flaws that elude single-paradigm approaches. The significant increase in detected vulnerabilities demonstrates the power of cognitive synergy facilitated by structured debate.

Nevertheless, the analysis also reveals avenues for future enhancement. The Deductive and Inductive agents alone identified 25 and 12 vulnerabilities, respectively, which were not part of \toolname's final verdict. This indicates that the current debate process may sometimes over-prioritize certain lines of reasoning, potentially missing some valid rule-based or pattern-based vulnerabilities. Future work could aim to refine the debate dynamics to ensure a more balanced synthesis of all expert insights, further strengthening the framework's detection capabilities.

\subsection{Threats to Validity}
\textbf{External Threats.} Our study faces several external threats. First, the choice of dataset language may introduce selection bias. \toolname performs excellently on C/C++ datasets, but its effectiveness in other programming languages (such as Python, Java, or JavaScript) has yet to be validated. Although our multi-agent debate framework is language-agnostic, the prompts and reasoning patterns for different languages may need adjustments. Future work will evaluate cross-language generalizability.
Secondly, there may be selection bias in the LLMs used in both \toolname and the baselines. Our experiments selected seven mainstream LLMs due to their strong performance on code-related tasks and their availability for research~\cite{he2025llm,zhu2025misum}. However, the rapid development of LLMs may lead to new models performing differently in vulnerability detection, limiting the applicability of our findings to models that were not evaluated. We mitigate this by selecting code-proficient models, with plans for broader evaluations.

\textbf{Internal Threats.} 
One potential threat is the accuracy of the baseline methods we have re-implemented. Differences in implementation details (e.g., hyperparameter settings) could affect performance comparisons. To mitigate this, we use official code and original configurations to ensure fairness. Another internal threat arises from the inherent randomness in the outputs of LLMs. To address this, we set the temperature to 0 to make the results more deterministic. Finally, the selection of optimal LLMs for each agent based on the PairVul dataset poses a potential risk of data leakage regarding the PrimeVul test set. To address this, we analyzed the data distribution differences between the two datasets. The PrimeVul covers more than 100 CWE types drawn from hundreds of large projects, whereas PairVul is mainly restricted to 5 CWE types from the Linux kernel. The calculated token-level Jensen–Shannon divergence between the two datasets is 0.43, indicating a substantial distributional shift. This quantitative evidence confirms that using PairVul for model selection does not leak information from the PrimeVul test set, ensuring the validity of our evaluation.


\section{Conclusion}
In this paper, we presented \toolname, a multi-perspective reasoning framework that advances automated vulnerability detection by integrating deductive, inductive, and abductive reasoning paradigms within a collaborative multi-agent system. Our approach achieves cognitive synergy through two stages, i.e., independent reasoning and collaborative debate, to resolve conflicts and identify subtle gaps that are overlooked by single-paradigm methods. Our experiments demonstrate that \toolname significantly outperforms state-of-the-art vulnerability detection methods, achieving a PairAcc of 40.00\% and an F1-score of 72.52\% on the PrimeVul dataset.
This research highlights the critical importance of cognitive diversity and structured debate in vulnerability detection, offering a promising approach to improving the security and reliability of software systems across diverse real-world scenarios.

\bibliographystyle{ACM-Reference-Format}
\bibliography{cite}

\end{document}